\documentclass[prl,reprint,amsart]{revtex4-1}

\usepackage{graphicx}
\usepackage{amssymb}
\usepackage{epstopdf}
\usepackage{mathrsfs}
\usepackage{amsmath}
\usepackage{verbatim}
\usepackage{color}
\usepackage{subfigure}
\usepackage{hyperref}

\definecolor{ORANGE}{rgb}{1.0,0.647,0.0} 

\begin{document}

\title{Multistep kinetic self-assembly of DNA-coated colloids}


\author{Lorenzo di Michele$^1$}
\author{Francesco Varrato$^2$}
\author{Jurij Kotar$^1$}
\author{Simon H. Nathan$^1$}
\author{Giuseppe Foffi$^{3,2}$}
\author{Erika Eiser$^1$}

\email[]{ee247@cam.ac.uk}
\affiliation{$^1$University of Cambridge, Cavendish Laboratory, JJ Thomson Avenue, Cambridge, CB3 0HE, United Kingdom}
\affiliation{$^2$Institute of Theoretical Physics, Ecole Polytechnique Federale de Lausanne, 1015 Lausanne, Switzerland}
\affiliation{$^3$Laboratoire de Physique de Solides, UMR 8502, B\^at. 510, Universit\'e Paris-Sud, F-91405 Orsay, France.}

\date{\today}

\maketitle

\textbf{Self-assembly is traditionally described as the process through which an initially disordered system relaxes towards an equilibrium ordered phase only driven by local interactions between its building blocks. However, This definition is too restrictive. Nature itself provides examples of amorphous, yet functional, materials assembled upon kinetically arresting the pathway towards the ground state \cite{Dong_2011,Yin_PNAS_2012,Dufresne_SoftMatter_2009,Stavenga_2011,Vukusic_Nature_2003}. \\
Kinetic self-assembly is intrinsically more flexible and reliable than its equilibrium counterpart, allowing control over the morphology of the final phase by tuning both the interactions and the thermodynamic pathway leading to kinetic arrest \cite{Grzybowski_SoftMtter_2009}.\\
Here we propose strategies to direct the gelation of two-component colloidal mixtures by sequentially activating selective interspecies and intra-species interactions. We investigate morphological changes in the structure of the arrested phases by means of event driven molecular dynamics (MD) simulations and experimentally using DNA-coated colloids (DNACCs). Our approach can be exploited to finely tune the morphology of multicomponent nano- or micro-porous materials with possible applications in hybrid photovoltaics, photonics and drug delivery.}\\

Thanks to the specificity of Watson-Crick base paring, DNACCs are considered as an ideal test bench for equilibrium self-assembly. 
Due to the strength and the sharp thermal activation of the DNA-mediated interactions \cite{Chaikin_PRL_2009,Crocker_PRL_2005,Crocker_PNAS_2012}, DNACCs reveal a strong tendency to kinetically arrest into amorphous phases, precursors of crystalline ground-states. Crystallization is only reliably achieved using nanoscale particles, \cite{Shultz_Nature_1996,Mirkin_Nature_1996,Mirkin_Nature_2008,Mirkin_Science_2011,Park_NMat_2010,Redl_Nature_2003,Gang_Nature_2008} for which interactions are weaker and their thermal activation smoother, whereas scaling up to slightly larger building blocks turns out to be difficult \cite{Angioletti-Uberti_NMat_2012,Crocker_Langmuir_2006}.\\
The properties that make DNACCs unsuitable for crystallization experiments make them ideal for the purpose of controlling their kinetic arrest in amorphous structures. We exploit both the selectivity and the sharp thermal activation of hybridization interactions to design binary mixtures of colloids in which interactions are activated at different stages of the quenching procedure. One of the species aggregates before the other, providing a template or a scaffold for the aggregation of the second.\\
In contrast to equilibrium self-assembly of DNACCs, only possible for nanoscale objects, our approach can be applied to the whole range of colloidal sizes.\\
In this letter we discuss two schemes that exemplify the potentiality of our strategy. Similar criteria can be applied to design arbitrarily complex systems with more than two components, multiple aggregation stages or modified interaction schemes.\\ 
In the final section, we describe some of the expected technological applications.\\

\begin{figure*}[ht!]
\includegraphics[width=17cm]{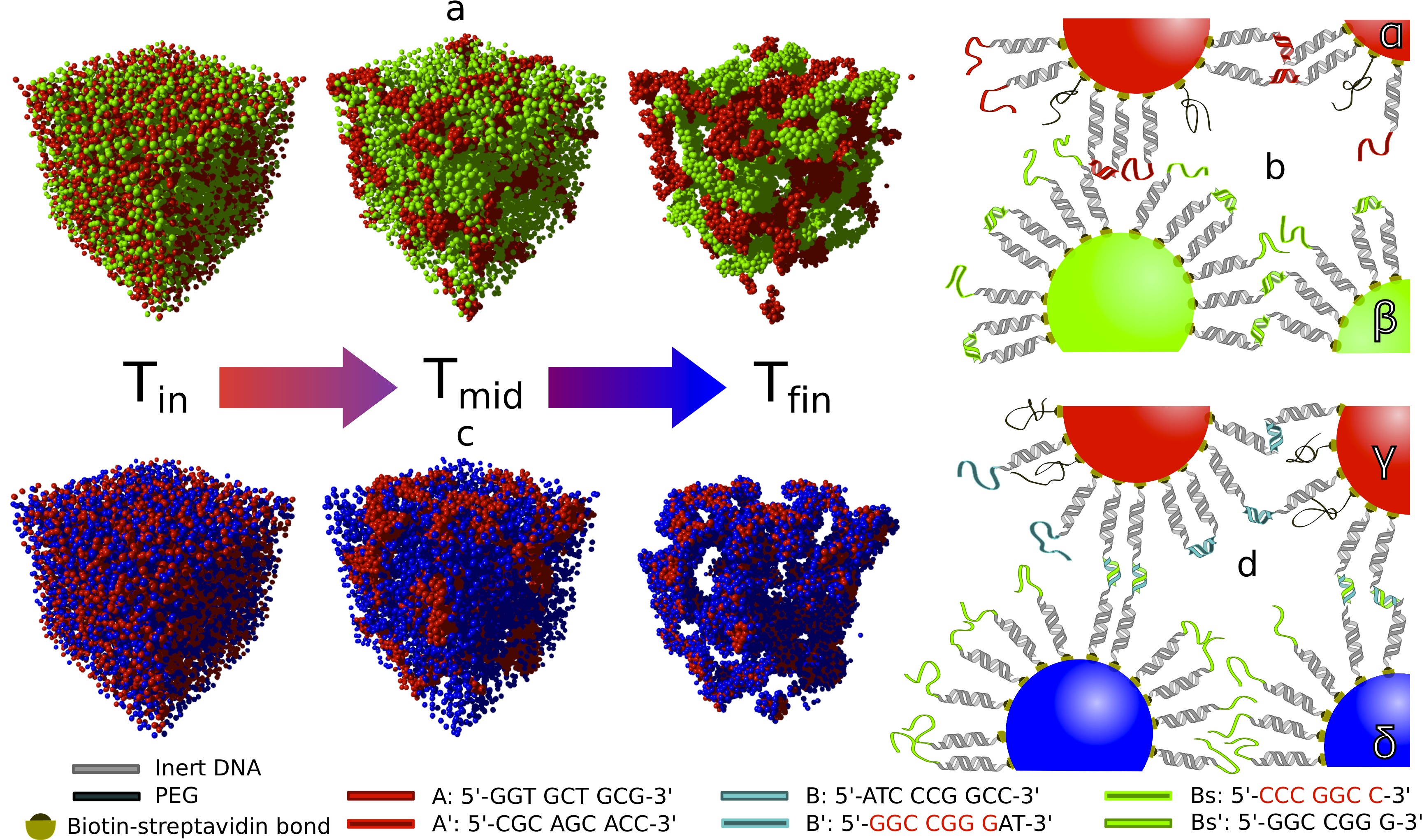}%
\caption{\label{fig1} \textbf{Self-assembly pathway and experimental implementation of the two binary mixtures.} \textbf{a}, From left to right, snapshots from MD simulations of the two-step gelation process of $\alpha-\beta$ mixture. Initially, colloidal interactions are repulsive and the system equilibrates into a uniform gas phase. The $\alpha-\alpha$ attraction is switched on, and gelation occurs. Finally the $\beta-\beta$ attraction is activated and the $\beta$ phase aggregates in the confined environment imposed by the $\alpha$ template-gel. \textbf{b}, Experimental implementation of the $\alpha$ and $\beta$ particles by means of DNACCs. $\alpha-\alpha$ and $\beta-\beta$ hybridization interactions are activated at temperatures $T_\alpha > T_\beta$ such that the two-step aggregation pathway can be induced upon quenching from $T_\mathrm{in}>T_\alpha$ to $T_\alpha>T_\mathrm{mid}>T_\beta$ and finally to $T_\mathrm{fin}<T_\beta$. \textbf{c}, From left to right, snapshots from MD simulations of a core-shell gelation of $\gamma-\delta$ mixture. Initially the $\gamma-\gamma$ attraction is switched on, and $\gamma$ particles form a gel from the initially uniform gas phase. Upon activation of the $\gamma-\delta$ attraction, $\delta$ particles coat the $\gamma$ gel as a single layer. $\gamma-\gamma$ interactions are repulsive at every stage. \textbf{d}, Experimental implementation of the $\gamma-\delta$ mixture with DNACCs. 
Colloids $\gamma$  are functionalised with $B$ and $B'$ sticky ends, which produce interspecies attraction for $T<T_\gamma$. Colloids $\delta$ are coated only with strand $B_\mathrm{S}$, complementary to part of $B'$ (marked in red), such that a $\gamma-\delta$ attraction arises for $T<T_\delta<T_\gamma$. The DNA sequences are grafted to the surface of the colloids via biotin-streptavidin bonds. To reduce the aggregation temperature of species $\alpha$ and $\gamma$ a fraction of the grafting sites are occupied by inert polyethylene-glycol (PEG) polymers.}
\end{figure*}

The first binary system is composed of species $\alpha$ and $\beta$. The pathway to kinetic arrest is summarized in Figure~\ref{fig1}~\textbf{a} as obtained with MD simulations. Initially the two-component mixture is equilibrated at high temperature, where it behaves as a one-component gas of hard spheres. A quench is then performed by suddenly lowering the temperature. At this stage, a short-range square-well attraction dominates between $\alpha$ colloids, which form a gel coexisting with the $\beta$ gas. Finally, an identical square-well attraction is activated between $\beta$ colloids, which gel in the confined environment imposed by the preexisting $\alpha$ network.\\

\begin{figure*}[ht!]
\includegraphics[width=17cm]{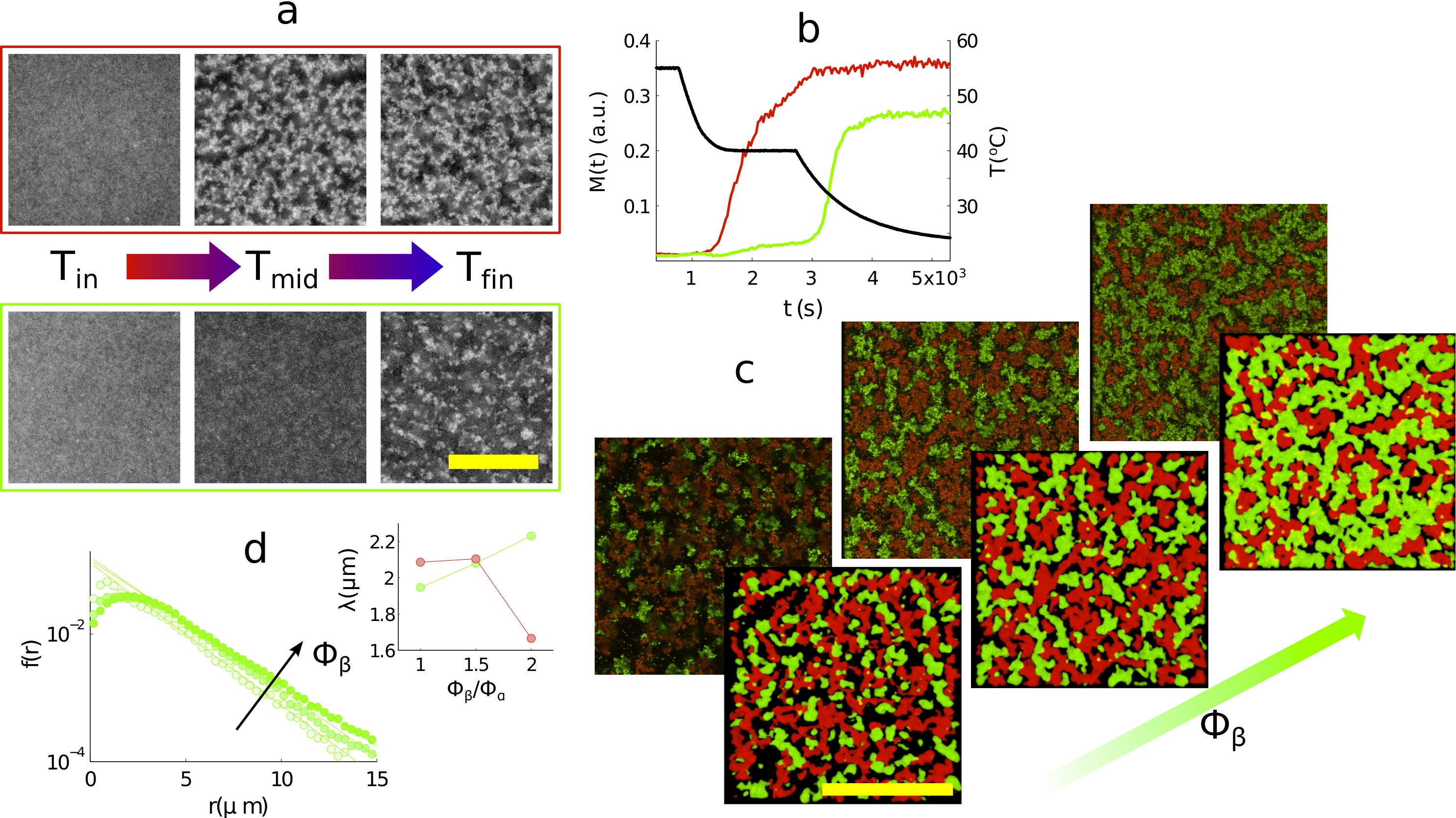}%
\caption{\label{fig2} \textbf{Two-step gelation with DNA-coated colloids}. \textbf{a}, Epifluorescence snapshots of $\alpha$ (top, red frame) and $\beta$ (bottom, green frame) colloids at different stages of the quenching procedure for a sample with $\phi_\alpha=\phi_\beta\approx 5\%$. The scale bar is equal to 20 $\mu$m. \textbf{b}, Evolution of temperature (right axis) and  $M(t)=\max\left[S_q(t)/S_q(t=0)\right]$ for the two components (left axis) for a sample with $\phi_\alpha\approx 5\%$ and $\phi_\beta\approx10\%$. \textbf{c}, Confocal fluorescence images of three arrested samples and corresponding top-view of 3D volume reconstructions with $\phi_{\alpha}\approx 5\%$ and, from left to right, $\phi_\beta\approx5,~7.5,~10\%$. The scale bar is equal to 50 $\mu$m. \textbf{d}, Chord distributions for the $\beta$ phase of samples shown in panel \textbf{c} ($\circ$) and exponential fits (solid lines). In the inset: persistence  length $\lambda$ evaluated from chord distributions of $\alpha$ and $\beta$ components for samples shown in panel \textbf{c}.}
\end{figure*}

The two-step kinetic arrest designed \emph{in silico} is tested experimentally using a binary mixture of fluorescently labeled DNA-coated microspheres with diameter of 500~nm.\\
As sketched in Figure~\ref{fig1}~\textbf{b}, colloids $\alpha$ are functionalised with a symmetric mixture (ratio 1:1) of double-stranded DNA (dsDNA) sequences terminated with single-stranded DNA (ssDNA) \emph{sticky ends} $A$ and $A'$, complementary to each other.  The use of dsDNA rigid spacers allows the sticky ends to explore large volumes around the grafting site and limits the entropic cost for hybridization \cite{Chaikin_PRL_2009,Angioletti-Uberti_NMat_2012,Crocker_PRL_2005,Crocker_PNAS_2012,Leunissen_NMat_2009,Leunissen_SoftMatter_2009,Leunissen_JACS_2010,Varrato_PNAS_2012}. Details about the preparation of DNA coatings are provided in the Supplementary Information (SI). Complementarity between ssDNA grafted on the same particle results in the formation of intra-colloid  \emph{loops} and inter-colloid \emph{bridges} \cite{Leunissen_NMat_2009,Leunissen_SoftMatter_2009,Leunissen_JACS_2010,Angioletti-Uberti_NMat_2012}. The latter are responsible for interspecies attraction between $\alpha$ colloids, activated by quenching below the aggregation temperature $T_\alpha\approx 43^\circ$C.\\ 
Similarly to $\alpha$, colloids $\beta$ are coated with a symmetric mixture of complementary $B_\mathrm{S}$ and $B'_\mathrm{S}$ strands. The aggregation between $\beta$ particles is however triggered at temperature $T_\beta\approx33^\circ$C$<T_\alpha$.\\
The two-step gelation is replicated by quenching the $\alpha-\beta$ mixture from the initial gas phase, allowed to equilibrate at $T_\mathrm{in}=55^\circ$C, to an intermediate temperature $T_\mathrm{mid}=40^\circ$C, with $T_\beta<T_\mathrm{mid}<T_\alpha$. At this stage, $\alpha$ colloids undergo gelation, whereas $\beta$ colloids remain in a gas phase. After a percolating $\alpha$ network is formed, we further decrease the temperature to $T_\mathrm{fin}<T_\beta$ to induce gelation of the $\beta$ species. Epifluorescence microscopy snapshots of the three stages are shown in Figure~\ref{fig2}~\textbf{a}.\\
We construct samples in a slab geometry to ensure high quality imaging, details are provided in the Methods section and SI.\\

We monitor the evolution of both the species by measuring the quantity $M(t)=\max\left[S_q(t)/S_q(t=0)\right]$, where $S_q(t)=\langle I_\textbf{q} I_{-\textbf{q}} \rangle / \int I(x,y) \mathrm{d}x\mathrm{d}y$. The intensity $I_\textbf{q}$ is the Fourier transform of an epifluorescence snapshot $I(x,y)$ and $\langle \dots \rangle$ indicates a radial average. The value $S_q(t=0)$ is measured for the gas phase at $T=T_\mathrm{in}$. The evolution of $M(t)$ for the $\alpha-\beta$ mixture, shown in Figure~\ref{fig2}~\textbf{b}, depicts the two-step gelation.\\
We obtain morphological information about the arrested structures from confocal imaging. We produce 3D reconstructions by processing confocal stacks with Gaussian filtering and thresholding algorithms (SI).  Quantitative characterisation is carried out by means of a chord-distribution analysis \cite{Testard_PRL_2011,Levitz_1998}. As shown in Figure S4, the chords are measured as the lengths of the intersections between straight lines drawn on the thresholded images and regions belonging to the aggregates. Chord length histograms $f(r)$ exhibit the exponential decay typical of patterns originating from arrested phase separation. The decay length $\lambda$ is defined as the persistence length of the structures \cite{Levitz_1998} and provides a measure of the length-scale of the kinetic arrest \cite{Testard_PRL_2011}.\\
In Figure~\ref{fig2}~\textbf{c}, we show coexisting $\alpha-\beta$ gels with a constant volume fraction for the $\alpha$ phase $\phi_\alpha\approx5\%$ and increasing $\phi_\beta\approx5,~7.5,~10\%$.  The morphology of the $\beta$ phase is influenced by the preexisting $\alpha$ gel. This is evident in the case of symmetric composition $\phi_\alpha=\phi_\beta$, for which, due to the confinement imposed by the $\alpha$ gel, the $\beta$ species aggregates into disconnected clusters. Such morphological asymmetry is confirmed by the chord-distribution length-scales $\lambda_\alpha>\lambda_\beta$ shown in the inset of Figure~\ref{fig2}~\textbf{d} and is a direct consequence of the two-step kinetic arrest. In fact,  $\alpha-\beta$ mixtures with $T_\alpha=T_\beta$ form coexisting gels with identical morphology \cite{Varrato_PNAS_2012} (see SI and Figure S5).
$\lambda_\beta$ grows upon increasing $\phi_\beta$ to 7.5 and 10\%. For the sample with $\phi_\beta=2\phi_\alpha$ we observe a significant drop in $\lambda_\alpha$, indicating that a high concentration of $\beta$ colloids in the gas phase inhibits the gelation of the $\alpha$ component.\\
Besides the confining effect of the $\alpha$ gel, the morphology of the $\beta$ phase can be influenced by nonspecific $\alpha-\beta$ attractions that might result in the trapping of $\beta$ particles within the $\alpha$ aggregates. By introducing controlled interspecies interactions mediated by DNA, we could take advantage of this effect and further manipulate the structures.\\
\begin{figure*}[ht!]
\includegraphics[width=17cm]{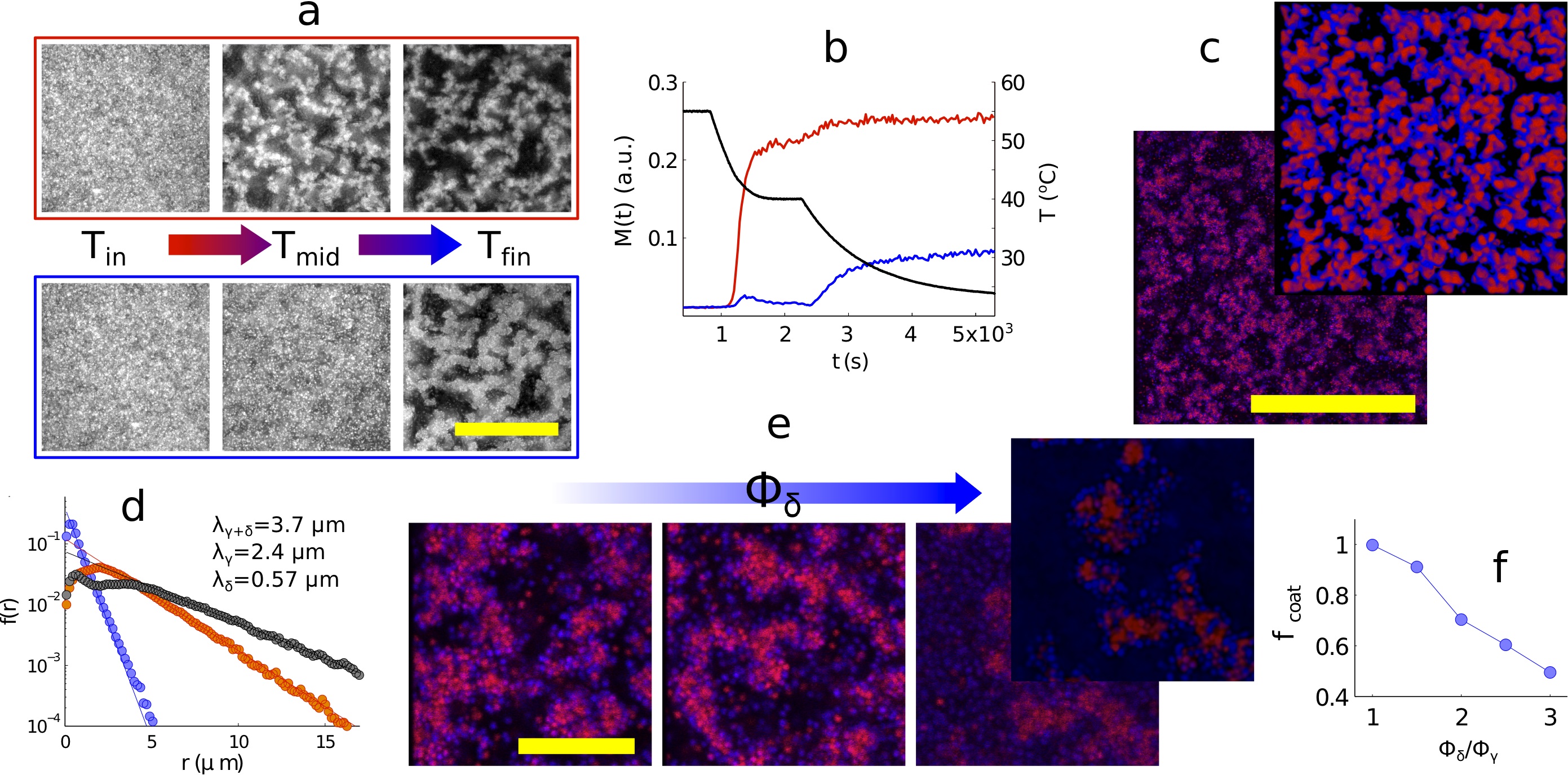}%
\caption{\label{fig3} \textbf{Core-shell gels with DNA-coated colloids}. \textbf{a}, Epifluorescence snapshots of $\gamma$ (top, red frame) and $\beta$ (bottom, blue frame) colloids at different stages of the quenching procedure for a sample with $\phi_\gamma=\phi_\delta\approx 5\%$. The scale bar is equal to 10 $\mu$m. In Supplementary Video 2 we show the evolution of the two-step gelation for a sample with $\phi_\gamma\approx5\%$ and $\phi_\delta\approx7.5\%$. \textbf{b}, Evolution of temperature (right axis) and of $M(t)=\max\left[S_q(t)/S_q(t=0)\right]$ for the two components (left axis) for a sample with  $\phi_\gamma=\phi_\delta\approx 5\%$. \textbf{c}, Confocal image and corresponding 3D reconstruction for a sample with  $\phi_\gamma=\phi_\delta\approx 5\%$. The scale bar is equal to $50 \mu$m. \textbf{d}, Chord distribution calculated for the sample in panel \textbf{c}. Blue $\circ$ indicate the $\delta$ phase,  red $\circ$ the $\gamma$ phase and  grey $\circ$ the total network obtained by adding up the $\delta$ and $\gamma$ images. The lines are exponential fits. \textbf{e}, Higher magnification confocal images showing samples with constant $\phi_\gamma=5\%$ and, from left to right, $\phi_\delta=5,~7.5,~10\%$. For the last sample we show a time-averaged image to distinguish the $\delta$ gas phase from the $\delta$ shell surrounding the $\gamma$ core gel. \textbf{f}, Fraction $f_\mathrm{coat}$ of $\delta$ colloids adhering to the $\gamma$ gel as a function of $\phi_\delta/\phi_\gamma$ calculated with MD simulations.}
\end{figure*}

This kinetic route can be exploited to produce structures characterised by an even higher level of complexity. We exemplify this by presenting a second binary mixture composed of colloids $\gamma$ and $\delta$.  The two-step arresting sequence is illustrated in Figure~\ref{fig1}~\textbf{c} as obtained from MD simulations. After equilibrating the system at high temperature, we trigger the gelation of the $\gamma$ component by activating the $\gamma-\gamma$ attraction, analogous to the first gelation step of the $\alpha-\beta$ mixture. For the second step, we activate an interspecies $\gamma-\delta$ attraction, maintaining the $\delta-\delta$ interaction repulsive. This results in the formation of a single layer of $\delta$ particles coating the percolating $\gamma$ gel, and producing a \textit{core-shell} gel structure.\\

Similarly to $\alpha$, $\gamma$ DNACCs are functionalised with a symmetric mixture of complementary sticky ends $B$ and $B'$, which cause aggregation at temperature $T<T_\gamma\approx43^\circ$C. Colloids $\delta$ are functionalised only with $B_\mathrm{S}$ sticky ends, complementary to a subsection of $B'$ (Figure~\ref{fig1}~\textbf{d}).  Upon quenching to temperatures below $T_\delta\approx33^\circ$C a specific $\gamma-\delta$ attraction is activated, which triggers the formation of the coating. Since $B_\mathrm{S}$ is not self-complementary, the $\delta-\delta$ interaction remains repulsive regardless of temperature.\\

The two-step formation of core-shell gels is demonstrated by epifluorescence images in Figure~\ref{fig3}~\textbf{a} and by $M(t)$ in Figure~\ref{fig3}~\textbf{b}, both relative to a sample with $\phi_\gamma=\phi_\delta\approx5\%$. In Figure~\ref{fig3}~\textbf{c} we show a confocal image with its 3D reconstruction. A and a zoomed confocal stack is shown in Supplementary Video 1.  Also for the core-shell structures, the chord distribution provides quantitative morphological information. We find a persistence length of the coating phase $\lambda_\delta\approx 570$~nm, as expected, similar to the diameter of a single colloid.  The persistence length of the total gels $\lambda_{\gamma+\delta}$ is consistently larger than $\lambda_\gamma$.\\
To further demonstrate that the $\delta-\delta$ attraction is negligible and $\delta$ colloids only form a single layer coating the surface of $\gamma$ gel, we produce samples with increasing $\phi_\delta$ and constant $\phi_\gamma\approx5\%$. High magnification confocal images are shown in Figure~\ref{fig3}~\textbf{e}, full sized ones in Figure S6. For the case of $\phi_\delta=5\%$, the amount of $\delta$ colloids in the gas phase is similar to the amount of $\gamma$ colloids. Both are compatible with a gas-gel coexistence. Increasing $\phi_\gamma$ to $\approx7.5\%$, we observe a clear growth of the number of free $\delta$ colloids, indicating that the surface of the $\gamma$ gel is saturated and further adhesion of the $\delta$ particles is hindered. This effect is more evident for samples with $\phi_\delta=10\%$, for which a time-averaged image is necessary to distinguish between the immobilized $\delta$ particles making up the shell and the abundant $\delta$ gas.\\ We quantify the saturation effect with MD simulations by measuring the fraction $f_\mathrm{coat}$ of $\delta$ colloids adhering to the $\gamma$ gel as a function of $\phi_\delta/\phi_\gamma$, as shown in Figure~\ref{fig3}~\textbf{f}. In agreement with 
 the experiments, the saturation is observed to occur for $1<\phi_\delta/\phi_\gamma<1.25$.\\


Multi-step kinetic self-assembly enables further morphological manipulation by changing the timescale of the quench process.
We can process the same $\alpha-\beta$ and $\gamma-\delta$ mixtures with a very fast, one-step, quench instead of the two-step one such that the second component is activated before the gelation of the first. For the $\alpha-\beta$ mixture, this removes the morphological asymmetry, while the $\gamma-\delta$ mixture aggregates in a single binary gel. All the morphologies between this limiting situation and the ones obtained the with two-step quench are in principle accessible and will originate materials with different tunable properties.\\

Colloidal structures assembled via DNA can be made permanent \cite{Leunissen_SoftMatter_2009}, enabling a direct application of our strategy to the self-assembly of functional materials.\\
For example, spongy phases generated in nature by arrested phase separation, can produce structural coloring  \cite{Stavenga_2011,Vukusic_Nature_2003,Yin_PNAS_2012,Dufresne_SoftMatter_2009}. Similar materials can be engineered based on our approach, with the additional possibility of tuning refractive index contrast by having multiple components.\\ In random lasers,  amorphous assemblies of colloids are immersed in a medium that provides optical gain to ensure the multiple scattering required to reach the lasing threshold \cite{Wiersma_NPhys_2008}. Resonant emission can be obtained by using randomly packed monodisperse spheres as scatterers \cite{Gottardo_NPhot_2008}. Similarly, we suggest the use of colloidal gels with tunable photonic response. Multiple components would also allow decoupling between the spatial distribution of the dyes providing gain, which can be incorporated in latex colloids \cite{Cerdan_NPhot_2012}, and the scatterers.\\
In hybrid photovoltaics, controlled phase separation is required between the conjugate polymer acting as light absorber/donor and inorganic particles acting as acceptors \cite{Bach_Nature_1998}. Schemes analogous to our core-shell self-assembly could be exploited to build inorganic semiconductor scaffolds of different morphology and coat them with organic particles.\\
Furthermore, the reversibility of the DNA interactions can be exploited to engineer responsive materials for sensing and drug release. Let us consider the core-shell gels. The scaffold phase can be made permanent and used as a means of containing and delivering controlled quantities of the coating colloidal species. The latter can then be released by an external stimulus, such as change in temperature or pH.\\
We conclude by stressing that our strategy is not limited to the use of DNA. Supramolecular chemistry provides a wide range of tags with selective interactions \cite{Taylor_ACSnano_2011}, the functionality of which is not limited to physiological environments as for DNA.\\

Summarizing, we introduce an alternative means of controlling the self-assembly of colloidal particles into amorphous materials.
Contrary to equilibrium self-assembly strategies, in which the target configuration only depends on the interactions between the building blocks, our approach relies on both the selectivity and the temperature dependence of the base-paring interactions and exploits multi-step kinetic arrest to tune the morphology of the final configuration.\\
We describe two model systems to demonstrate the potential and flexibility of our scheme. The first is a binary mixture undergoing a two-step gelation, with the second aggregating phase forced to gel in the confined environment produced by the first aggregating phase. The second system self-assembles in core-shell colloidal gels, with the second component coating the  network firstly formed, which acts as a scaffold.\\
Multi-step kinetic self-assembly of mesoscopic building blocks mediated by DNA specific interactions holds great promise as general approach for reliable self-assembly of mesoporous materials.

\subsection*{Methods}
\subsubsection*{Experimental methods}
{ \footnotesize DNA (Integrated DNA technology) is grafted to the surface of streptavidin-coated colloids via biotin-streptavidin linkage. The DNA constructs are structured as follows: Biotin--5'--TTTTT--dsDNA spacer--TTTTT-ssDNA sticky end--3'. The dsDNA spacer has a length of 60 base pairs. The 5-thymine sequences are added to confer more flexibility to the constructs. The melting temperatures of the free sticky ends in solution are: $B-B'$: 37.5$^\circ$C, $A-A'$: 37.8$^\circ$C, $B_\mathrm{S}-B_\mathrm{S}' (A')$: 30.2$^\circ$C, calculated at 50mM ionic strength. We use two batches of green and red fluorescent polystyrene colloids (Microparticles GmbH, Berlin). The experiments are carried out in 10-mM Tris-HCl pH 8 + 1-mM EDTA buffer with the addition of 50-mM NaCl. The density of the solution is raised to 1.05 g/cm$^3$ to match that of polystyrene. Due to the high refractive index of polystyrene and the incompatibility of DNA with organic solvents, refractive index matching limitations exist. To allow good quality optical imaging of opaque samples, glass sample chambers are built in a slab geometry, with width of 1.8 cm and thickness of ~15 $\mu$m. The chambers are permanently sealed to avoid evaporation. Further details are provided in SI. Heating/quenching procedures are carried out while imaging with a Nikon Eclipse Ti-E inverted microscope using Plan Apo VC 20$\times$/0.75 or Plan Fluor 40$\times$/0.75 dry objectives. Confocal imaging is performed on a Leica TCS SP5 microscope equipped with an HCX PL APO CS 100$\times$ 1.4 oil immersion objective.}
\subsubsection*{Computer simulations}
{\footnotesize Event-driven molecular dynamics simulations are performed in cubic boxes with periodic boundary conditions. The colloids are modeled by means of a square-well potential, with a hard-sphere core and with a well whose depth is either set equal to $\epsilon_{ij}=-\epsilon_0$ when the $i$ and $j$ species are attractive, or to $\epsilon_{ij}=0$ when they are considered as repulsive. The simulations involve up to $2\times10^4$ particles. The colloids have equal diameter $\sigma=1$ and equal mass $m=1$. The potential is chosen to be short-ranged, i.e. the $3\%$ of the diameter. We used reduced units, $k_B=1$ and $\epsilon_0=1$. The quenches are performed by lowering the temperature from $T=100$ to the final value $T=0.05$, well below the known critical temperature $T_c=0.3$ for a single-component system \cite{Miller_JPhysCond_2004}. As the system falls out of equilibrium, averages over 10 independent realizations of the initial conditions are considered instead of time averages.}
\subsection*{Acknowledgments}
{F.V. and G.F. acknowledge financial support from Swiss National Science Foundation (SNSF) Grants PP0022\_119006 and PP00P2\_140822/1. L.D. acknowledges financial support from Marie Curie Initial Training Network Grant ITN-COMPLOIDS 234810.

\end{document}